\documentclass[nofootinbib]{revtex4}
\usepackage{graphicx}
\usepackage{latexsym}
\def\be{\begin{equation}}
\def\ee{\end{equation}}
\def\bea{\begin{eqnarray}}
\def\eea{\end{eqnarray}}
\def\bi#1{\hbox{\boldmath{$#1$}}}
\newcommand{\wjma}[6]{\left(
                           \begin{array}{ccc}
         #1 & #2  & #3  \\
         #4 & #5  & #6
                           \end{array}
                   \right)}

\begin{document}

\title{Spacetime variation of $\alpha$ and the CMB power spectra after the recombination}

\author{Xiulian Wang}
\author{Mingzhe Li}
\email{limz@nju.edu.cn}
\affiliation{ Department of Physics, Nanjing
University, Nanjing 210093, People's Republic of China}
\affiliation{Joint Center for Particle, Nuclear Physics and
Cosmology, Nanjing University - Purple Mountain Observatory, Nanjing
210093, Peoples Republic of China}


\begin{abstract}
The possible variation of the fine structure constant may be due to the non-minimal coupling of the electromagnetic field to a
light scalar field which can be the candidate of dark energy. Its dynamical nature renders the fine structure constant
varies with time as well as space. In this paper we point out the spatial fluctuation of the fine structure will modify the power spectra of
the temperature and the polarization of the cosmic microwave background. We show explicitly that the fluctuations
of the coupled scalar field generate new temperature anisotropies at the linear order and induce a $B$ mode to the polarization at higher order
in general.
\end{abstract}

\maketitle

\hskip 1.6cm PACS number(s): 98.70.Vc \vskip 0.4cm

\section{Introduction}

There is a long history to study the possible variations of fundamental constants \cite{history,history2,history3,history4}.
Simply these variations can be modeled as a light scalar field non-minimally coupled to the matter fields.
The claimed evidence \cite{webb,webb2} for a smaller fine structure constant $\alpha$ at the redshift of $z\sim 3$ from
quasar observations stimulated many interests in investigating different theories of varying $\alpha$ \cite{varying,varying2,varying3,
varying4,varying5,varying6,varying7,varying8,varying9,varying10,varying11,varying12,varying13,varying14,varying15,varying16,varying17,varying18,
varying19,varying20,varying21,varying22,varying23,varying24,varying25,varying26,varying27,varying28}.
On the other hand, the discovery of the accelerating expansion of the universe requires in general relativity the existence
of dark energy which, if dynamical, is extensively modeled as an ultra-light scalar field. Hence it is natural to
ask whether the variation of the fine structure constant is induced by the non-minimal coupling of dark energy to the
electromagnetic field. This connection provides another way to probe the dark energy in the Universe.

The variation of fine structure constant has effects on the cosmic microwave background radiation (CMB).
As discussed in Refs. \cite{Hannestad:1998xp,Kaplinghat:1998ry}, the change in $\alpha$ will change the history of
recombination and the position of the first Doppler peak of the CMB spectrum and
is constrained by the observational data \cite{Avelino:2000ea,Avelino:2001nr,Rocha:2003gc,Ichikawa:2006nm}.
If the variation of $\alpha$ in time is brought by the coupling of a light cosmic field $\phi$, there should be spatial variation
of $\alpha$ because of the dynamical nature of $\phi$. The effects of spatial
fluctuations of $\alpha$ on CMB are firstly studied in Ref. \cite{Sigurdson:2003pd} by considering the implicit dependence of
CMB temperature and polarization fields on $\alpha$ due to the recombination history. At different places, the recombination processes are different.
The authors showed that these spatial fluctuations induce
a $B$ mode to the polarization and non-Gaussian temperature and polarization correlations.
In this paper, we study this problem again focusing on the later time evolution after the recombination.
For simplicity we will not consider the reionization and other scattering effects.
We show explicitly the modifications to the power spectra without
focusing on a specific model of the coupled scalar field $\phi$. As we will point out,
the spatial fluctuations of $\phi$ generate new temperature anisotropies at the linear order. This
is similar to the Sachs-Wolf effect caused by the inhomogeneities of the gravitational field because
the coupling of $\phi$ induce an extra long range force between photons. So, photons from different positions on the last scattering
surface to us will lose or get different extra energies. Furthermore,
the polarization spectra and the temperature-polarization cross spectrum are also modified at higher order. In particular this provides a new mechanism
to produce CMB B-mode polarization analogous to weak lensing. Our study suggests in the future studies on varying $\alpha$ theories, especially
on the constraints on the models from CMB data, the effects from the spatial fluctuations of $\alpha$ should also be considered.

This paper is organized as follows. In Section II, we will discuss the transportation of CMB photons after the recombination based on
the method of geometric optics approximation. This method had been used in Ref. \cite{Li:2006ss} and
more systematically in \cite{Li:2008tma} to analyze the rotation of polarization when the photon coupled to an external field anomalously
\cite{Feng:2004mq,Feng:2006dp};
In Section III, we derive the modified power spectra. We calculate the temperature spectrum at the linear order and the polarization
spectra at higher order; Section IV is the conclusion.

\section{The transportation of CMB photon after recombination}

As mentioned above, the varying $\alpha$ theory maybe modeled as the electromagnetic field non-minimally coupled to a scalar field
which may or may not be the dark energy.
The Lagrangian for this modified electromagnetic theory can be written as,
\be\label{lagrangian}
\mathcal{L}=f(\phi)F_{\mu\nu}F^{\mu\nu}~.
\ee
Where $f(\phi)$ is a dimensionless function of $\phi$. Without the induced coupling, it is normalized as $-1/4$.
The fine structure constant is
$\alpha=-\alpha_0/4f$ with
$\alpha_0\sim 1/137$ being the fine structure constant at present time. The difference of $\alpha$ at the redshift $z$ from
today is
\be
\frac{\delta\alpha}{\alpha_0}=\frac{\alpha_z-\alpha_0}{\alpha_0}=-\frac{1+4f_z}{4f_z}~,
\ee
and $f_0$ should be equal to $-1/4$.
With the Lagrangian (\ref{lagrangian}) we obtain the equation of motion,
\be\label{xx}
\nabla_{\mu}F^{\mu\nu}= -\nabla_{\mu}\ln f
F^{\mu\nu}~.
\ee
In addition we have the
identity: \be
\nabla_{\mu}F_{\rho\sigma}+\nabla_{\rho}F_{\sigma\mu}+\nabla_{\sigma}F_{\mu\rho}=0~.\label{identity}
\ee
The transportation equations of CMB photons can be obtained in terms of the geometric optics approximation (GOA)
which has been used
to study the $CPT$ violating effects of the anomaly coupling of photons \cite{Li:2006ss,Li:2008tma}.
Similar to the analysis in Ref. \cite{Li:2008tma}, we first differentiate Eq.
(\ref{identity}) to get a second order differential equation for $F_{\mu\nu}$
\bea\label{master} \Box
F_{\rho\sigma}+\nabla_{\rho}(\nabla^{\mu}\ln f F_{\mu\sigma})
-\nabla_{\sigma}(\nabla^{\mu}\ln f F_{\mu\rho})
-[F^{\alpha}_{~\rho}R_{\alpha\sigma}-F^{\alpha}_{~\sigma}R_{\alpha\rho}-F^{\mu\alpha}R_{\alpha\mu\rho\sigma}]
=0~, \eea
where $R_{\alpha\sigma}$ and $R_{\alpha\mu\rho\sigma}$ are
Ricci and Riemann tensors respectively and will be neglected in GOA in the following discussions.
Then we make the ansatz
\be\label{goa} F^{\mu\nu}=(a^{\mu\nu}+\epsilon
b^{\mu\nu}+\epsilon^2 c^{\mu\nu}+...)e^{iS/\epsilon}~,
\ee where $\epsilon$
is a small real parameter and $S$ is a real function. We define the wave vector as \be k_{\mu}\equiv
\nabla_{\mu}S~, \ee which represents the travel direction of the
photon. Substituting the ansatz (\ref{goa}) into Eqs. (\ref{master})
and (\ref{identity}) we have the equations expanded by $\epsilon$
\bea\label{master2} &
&\Box(a_{\rho\sigma}+\epsilon
b_{\rho\sigma}+...)+\frac{2i}{\epsilon}k^{\mu}
\nabla_{\mu}(a_{\rho\sigma}+\epsilon b_{\rho\sigma}+...)
+\frac{i}{\epsilon}(\nabla_{\mu}k^{\mu}) (a_{\rho\sigma}+\epsilon
b_{\rho\sigma}+...)-\frac{1}{\epsilon^2}k_{\mu}k^{\mu}
(a_{\rho\sigma}+\epsilon b_{\rho\sigma}+...)\nonumber\\
&=&-[ (\nabla_{\rho}\nabla^{\mu}\ln f)(a_{\mu\sigma}+\epsilon b_{\mu\sigma}+...)
+\nabla^{\mu}\ln f(\nabla_{\rho}a_{\mu\sigma}+\epsilon \nabla_{\rho}b_{\mu\sigma}+...)
+\frac{ik_{\rho}}{\epsilon}\nabla^{\mu}\ln f(a_{\mu\sigma}
+\epsilon b_{\mu\sigma}+...)]+[\rho\leftrightarrow \sigma]
\eea
and
\be\label{identity2}
[\nabla_{\mu}(a_{\rho\sigma}+\epsilon b_{\rho\sigma}+...)+\frac{i}{\epsilon}k_{\mu}(a_{\rho\sigma}+
\epsilon b_{\rho\sigma}+...)]+[\rho\sigma\mu]+[\sigma\mu\rho]=0~.
\ee
At the leading order of the GOA, Eq. (\ref{identity2}) gives \be
k_{\mu}a_{\rho\sigma}+k_{\rho}a_{\sigma\mu}+k_{\sigma}a_{\mu\rho}=0
\ee which implies that $a_{\rho\sigma}$ should have the following
antisymmetric form: \be\label{solution1}
a_{\rho\sigma}=k_{\rho}a_{\sigma}-k_{\sigma}a_{\rho}~. \ee Then we
collect the terms of Eq. (\ref{master2}) at the orders
of $1/\epsilon^2$ and $1/\epsilon$, respectively. At the order of
$1/\epsilon^2$, we have \be\label{solution2}
k_{\mu}k^{\mu}=0~. \ee
This means the dispersion relation is unchanged at the most leading order of GOA.
The propagation equation of $k^{\mu}$ can be
obtained via differentiating the above equation again:
\be\label{solution21}
0=\nabla_{\nu}(k_{\mu}k^{\mu})=2\nabla^{\mu}S\nabla_{\nu}\nabla_{\mu}S=
2\nabla^{\mu}S\nabla_{\mu}\nabla_{\nu}S
=2k^{\mu}\nabla_{\mu}k_{\nu}~. \ee
This indicates $k^{\mu}$ is parallelly transported along the light curve, the null
geodesic. The
vector $k^{\mu}$ defines an affine parameter $\lambda$ which
measures the distance along the light ray: \be k^{\mu}\equiv
\frac{dx^{\mu}}{d\lambda}~. \ee
The transportation equations for the vector $a^{\nu}$ are obtained at the order of $1/\epsilon$: \be\label{solution3}
\mathcal{D}a^{\nu}+\frac{1}{2}(\theta+\mathcal{D}\ln f)a^{\nu}=0~, \ee where we
have considered Eq. (\ref{solution1}) and defined the operator
$ \mathcal{D}\equiv k^{\mu}\nabla_{\mu}=d/d\lambda$. The quantity $\theta=
\nabla_{\mu}k^{\mu}$ describes the expansion of the bundle of the
light. In addition, by applying the GOA to
Eq. (\ref{xx}), we have \be\label{solution4}
k_{\mu}a^{\mu}=0~. \ee The basic results we got above are
Eqs. (\ref{solution3}) and (\ref{solution21}) with two
orthogonality relations (\ref{solution2}) and (\ref{solution4}).

It is convenient to use the Stokes parameters to study the
polarization of radiation. The four Stokes parameters are well
defined in the local inertial frame (Lorentz frame). Considering a
monochromatic electromagnetic wave of frequency $\omega_0$
propagating in the $+z$ direction,
the Stokes parameters
are defined as the following time averages
\bea\label{stokes}
     I & \equiv & \left\langle E_x E_x^{\ast}\right\rangle
                         +\left\langle E_y E_y^{\ast}\right\rangle ,\nonumber\\
     Q & \equiv & \left\langle E_x E_x^{\ast}\right\rangle
                         -\left\langle E_y E_y^{\ast}\right\rangle ,\nonumber\\
     U & \equiv & \left\langle E_x E_y^{\ast}\right\rangle
                         +\left\langle E_x^{\ast}E_y\right\rangle ,\nonumber\\
     V & \equiv & i[\left\langle E_x E_y^{\ast}\right\rangle
                         -\left\langle E_x^{\ast} E_y\right\rangle ].
\eea
We can use the tetrad formalism to get these parameters in the coordinate frame.
A tetrad is a set of four
orthogonal unit basis vectors $e^{\mu}_{~(a)}$, with $a=0,1,2,3$.
The Latin indices are
lowered and raised by the Minkowski metric $\eta^{ab}$, the Greek
indices, however, by the coordinate metric $g^{\mu\nu}$.
The tetrad
has the following properties: \be
g_{\mu\nu}e^{\mu}_{~(a)}e^{\nu}_{~(b)}=\eta_{ab}~,~~~\eta^{ab}e^{\mu}_{~(a)}e^{\nu}_{~(b)}=g^{\mu\nu}~.
\ee
As discussed in \cite{Li:2008tma}, if we require the observer at rest with the local inertial frame at each point to
see the light traveling along the $+z$ direction, the zeroth and the third components of the tetrad vectors should be
\be
e^{\mu}_{~(0)}=u^{\mu}~,~~~e^{\mu}_{~(3)}=\frac{1}{\omega}(k^{\mu}-\omega
u^{\mu})~, \ee where $u^{\mu}$ is the four-velocity of the observer and $\omega\equiv k_{\mu}u^{\mu}$ is the frequency
measured by him (or her). The other tetrad vectors $e^{\mu}_{~(1)}$
and $e^{\mu}_{~(2)}$ are undetermined, but they should be unit spacelike, orthogonal to each other
and to $e^{\mu}_{~(0)}$, $e^{\mu}_{~(3)}$, and therefore orthogonal to
$k^{\mu}$.

The electric vector in general spacetime for the local observer is  \be E^{\mu}\equiv
F^{\mu\nu}u_{\nu}~. \ee At the leading order of the GOA, it is \be E^{\mu}=
a^{\mu\nu}u_{\nu}e^{iS/\epsilon}=(k^{\mu}a^{\nu}-k^{\nu}a^{\mu})u_{\nu}e^{iS/\epsilon}~.\ee
Transforming it to the local inertial frame, we get the $x$ and $y$
components of the electric field in this frame easily: \be
E_x=\bar{E}_1=E_{\mu} e^{\mu}_{~(1)}~,~~~E_y=\bar{E}_2=E_{\mu}
e^{\mu}_{~(2)}~. \ee With
above equations and Eqs. (\ref{stokes}), we have the expressions of
the Stokes parameters in the coordinate frame
\cite{anile,Kopeikin:2005jm}: \bea
I&=&\omega^2 L_{\mu\nu}(e^{\mu}_{~(1)} e^{\nu}_{~(1)}+e^{\mu}_{~(2)} e^{\nu}_{~(2)})\nonumber\\
Q&=&\omega^2 L_{\mu\nu}(e^{\mu}_{~(1)} e^{\nu}_{~(1)}-e^{\mu}_{~(2)} e^{\nu}_{~(2)})\nonumber\\
U&=&\omega^2 L_{\mu\nu}(e^{\mu}_{~(1)} e^{\nu}_{~(2)}+e^{\mu}_{~(2)} e^{\nu}_{~(1)})\nonumber\\
V&=&i\omega^2 L_{\mu\nu}(e^{\mu}_{~(1)}
e^{\nu}_{~(2)}-e^{\mu}_{~(2)} e^{\nu}_{~(1)})~, \eea where
$L_{\mu\nu}\equiv <a_{\mu}a_{\nu}^{\ast}>$ satisfies the following
equation making use of Eq. (\ref{solution3}): \be \mathcal{D}
L_{\mu\nu}+(\theta+\mathcal{D}\ln f)
L_{\mu\nu}=0~.
\ee We require the tetrad frames to be not physically
rotating. In order to do that, we set the tetrad vectors at each
point so that $e^{\mu}_{~(1)}$ and $e^{\mu}_{~(2)}$ are parallelly
transported along the light curve. So it is straightforward to get
the propagation equations of the four parameters along the light
curve: \be
\mathcal{D} F_a+(\theta+\mathcal{D}\ln f) F_a=0~,\ee where $F_a\equiv S_a/\omega^2\equiv
(I,~Q,~U,~V)/\omega^2$.
Hence the observed Stokes
parameters today should be
\be
S^{obs}_{a0}=-4\frac{\omega_0^2}{\omega_r^2}f_rS_{ar}\exp ({\int^{\lambda_r}_{\lambda_0}\theta d\lambda})~.
\ee
In above $r$ means the moment when the recombination is finished and we have used $f_0=-1/4$.
From above equation we can see that the effect of the coupling in (\ref{lagrangian}) on CMB at later time
after the recombination is just indicated by the ``dilation" factor $f_r$.
To disentangle this later time effect, we rewrite Eq. (\ref{ro}) as
\be\label{ro}
S^{obs}_{a0}= -4f_r S_a~,
\ee
where $S_{a}$ represent the should be Stokes parameters observed today without later time variation of $f$. But it includes
the effect from the modified recombination history.
This is the basic result obtained in
this section. The Stokes V cannot be generated by Thomson scattering and can be neglected.
Because $f_r$ is a function of the scalar field $\phi$ at the last scattering surface, it has different values at different positions.
So, it can be decomposed as $f_r=\bar{f}_r+\delta f_r$. The background part $\bar{f}_r$ differs from $-1/4$ because of the time variation.
It changes the redshift of the recombination slightly. In the next section we will consider the effects of spatial fluctuation $\delta f_r$
on CMB power spectra.

\section{CMB Power Spectra}

In the spatially flat Universe, we can expand the temperature and polarization
anisotropies in terms of appropriate spin-weighted spherical harmonic
functions on the sky \cite{Zaldarriaga:1996xe}: \bea
\Theta(\hat{\bi{n}})&=& \sum_{lm}a_{\Theta,lm}Y_{lm}(\hat{\bi{n}})\nonumber \\
(Q\pm iU) (\hat{\bi{n}})&=& \sum_{lm} a_{\pm 2, lm} \;_{\pm 2}Y_{lm}(\hat{\bi{n}})~.
\eea
Where $\Theta(\hat{\bi{n}})\equiv \Delta T(\hat{\bi{n}})/T$.
The expressions for the expansion coefficients are
\begin{eqnarray}
a_{\Theta,lm}&=&\int d\Omega\; Y_{lm}^{*}(\hat{\bi{n}}) \Theta(\hat{\bi{n}})
\nonumber  \\
a_{\pm 2,lm}&=&\int d\Omega \;_{\pm 2}Y_{lm}^{*}(\hat{\bi{n}}) (Q\pm iU)(\hat{\bi{n}})~.\label{alm}
\end{eqnarray}
Instead of $a_{2,lm}$ and $a_{-2,lm}$, it is convenient to introduce their
linear combinations
\begin{eqnarray}
a_{E,lm}=-(a_{2,lm}+a_{-2,lm})/2 \nonumber \\
a_{B,lm}=i(a_{2,lm}-a_{-2,lm})/2.
\label{aeb}
\end{eqnarray}
The power spectra are defined as \be \langle a_{X',l^\prime
m^\prime}^{*} a_{X,lm}\rangle= C^{X'X}_{l} \delta_{l^\prime l}
\delta_{m^\prime m} \ee with the assumption of statistical isotropy.
In the equation above, $X'$ and $X$ denote the temperature $\Theta$ and
the $E$ and $B$ modes of the polarization field, respectively. For Gaussian
theories, the statistical properties of the CMB
temperature/polarization map are specified fully by these
spectra. Without parity violation, $C^{\Theta B}_l=C^{EB}_l=0$.

Consider the modification in Eq. (\ref{ro}), the power spectra would
be changed. The temperature fluctuations are changed as \be
\Theta^{obs}=\frac{1}{4}\frac{\delta f_r}{\bar{f}_r}+(1+\frac{\delta
f_r}{\bar{f}_r})\Theta~. \ee This means the temperature fluctuations
receive a linear order modification. This is similar to Sachs-Wolf
effect of the gravitational field on CMB. The coupling of $\phi$ to
photons induce an extra long range force. The photons from different
positions on the last scattering surface to us will lose or get
different extra energies due to the inhomogeneous distribution of
$\phi$ on that surface. To evaluate the corrections, we expand $f_r$
by $\delta\phi_r$ to quadratical order, \be
f_r(\phi)=\bar{f}_r+A_1\delta\phi_r+A_2\delta\phi_r^2~. \ee We
assume $\delta\phi_r$ is a Gaussian random field, so it can be also
described by a power spectrum. Expand it on the sky \be\label{chi2}
\delta\phi_r=\sum_{lm}b_{lm}Y_{lm}(\hat{\bi{n}})~, \ee and define
its angular power spectrum as \be\label{chi3} \langle b_{l^\prime
m^\prime}^{*} b_{lm}\rangle= C^{\phi}_{l} \delta_{l^\prime l}
\delta_{m^\prime m} ~, \ee where we have assumed statistical
isotropy of $b_{lm}$. Furthermore, we have \be
C^{\phi}_l=4\pi \int \frac{dk}{k}\mathcal{P}_{\phi}(k) j_l^2(k\Delta \eta)~,~~~
\sum_{l}(2l+1)C^{\phi}_{l}=4\pi \langle\delta\phi_r^2\rangle~, \ee
where $\mathcal{P}_{\phi}(k)$ is the power spectrum of the perturbations of $\phi$ at the last scattering surface, $j_l$ is the
spherical Bessel function and
$\Delta\eta=\eta_0-\eta_r$ is conformal time difference between the time of last scattering and today.

In general, $\delta\phi_r$ has correlations with $\Theta$ and $E$.
In this paper, we will not consider this complication by assuming
$\delta\phi$ has different origin from that of density perturbation
of the photon. With this assumption the modified temperature power
spectrum would be \bea
C^{\Theta\Theta,obs}_l&=&C^{\Theta\Theta}_l+(\frac{A_1}{4\bar{f}_r})^2C^{\phi}_l+\frac{1}{32\pi}(\frac{A_2}{\bar{f_r}})^2\sum_{l_1l_2}
\wjma{l}{l_1}{l_2}{0}{0}{0}^2(2l_1+1)(2l_2+1)C^{\phi}_{l_1}C^{\phi}_{l_2}\nonumber\\
&+&\frac{1}{4\pi}(\frac{A_1}{\bar{f_r}})^2\sum_{l_1l_2}
\wjma{l}{l_1}{l_2}{0}{0}{0}^2(2l_1+1)(2l_2+1)C^{\phi}_{l_1}C^{\Theta\Theta}_{l_2}
~,
\eea
where $\wjma{l}{l_1}{l_2}{m}{m_1}{m_2}$ is the Wigner 3j symbol.

The polarizations of CMB are already at the perturbative level. They don't have background parts, so,
\be
(Q\pm iU)^{obs}=-4\bar{f}_r(1+\frac{\delta f_r}{\bar{f}_r})(Q\pm iU)~.
\ee
The modifications can only appear at higher orders.
The calculations of the polarization spectra and the cross spectrum of temperature and polarization are long.
Similar calculations can be found in Ref. \cite{Li:2008tma} for the modified spectra by anomaly coupling.
Here we only present the results up to the second order in the following
 \bea\label{rotationformulas2}
 C^{\Theta E,obs}_l&=&-4(\bar{f}_r+A_2\langle\delta\phi_r^2\rangle)C^{\Theta E}_l\nonumber\\
 &-&\frac{1}{2\pi}\frac{A_1^2}{\bar{f_r}}\sum_{l_1l_2}
\wjma{l}{l_1}{l_2}{0}{0}{0}\wjma{l}{l_1}{l_2}{2}{-2}{0}(1+(-1)^L)(2l_1+1)(2l_2+1)C^{\phi}_{l_1}C^{\Theta E}_{l_2}~,\nonumber\\
 C^{EE,obs}_l&=& 16(\bar{f}_r^2+2\bar{f}_rA_2\langle\delta\phi_r^2\rangle)C^{EE}_l\nonumber\\
 &+&\frac{2A_1^2}{\pi}\sum_{l_1l_2} \wjma{l}{l_1}{l_2}{2}{-2}{0}^2(2l_1+1)(2l_2+1)C^{\phi}_{l_2}\{[1+(-1)^{L}
 ]C^{EE}_{l_1} +[1+(-1)^{L+1}]C^{BB}_{l_1}\}~,\nonumber\\
  C^{BB,obs}_l&=& 16(\bar{f}_r^2+2\bar{g}_rA_2\langle\delta\phi_r^2\rangle)C^{BB}_l\nonumber\\
 &+&\frac{2A_1^2}{\pi}\sum_{l_1l_2} \wjma{l}{l_1}{l_2}{2}{-2}{0}^2(2l_1+1)(2l_2+1)C^{\phi}_{l_2}\{[1+(-1)^{L+1}
 ]C^{EE}_{l_1} +[1+(-1)^{L}]C^{BB}_{l_1}\}~,
  \eea
where $L=l+l_1+l_2$.
From these equations, we can see that the fluctuation of $\phi$ induce a mixing between $E$ and $B$ modes.
Even though the original $B$ mode generated by primordial gravitational wave from inflation is negligibly small,
it can still be produced from the $E$ mode and $\phi$ with odd $L$. These $B$ modes originated from the fluctuations of
fine structure are mixed with those from CMB weak lensing which is due to metric perturbations at the second order.
It is hard to distinguish between them. However, they can in principle be distinguished from the $B$ modes rotated from $E$ modes
by the anomaly coupling as studied in \cite{Li:2008tma}. As the third equation of \ref{rotationformulas2} showed, only the $B$ modes
with $L=odd$ are produced from $E$-mode polarizations. But for the case studied in \cite{Li:2008tma}, the primordial $E$-modes mainly
rotated to $B$ polarizations with $L=even$ (see the fifth equation of (69) in Ref. \cite{Li:2008tma}).

\section{Conclusion}

The time variation of the fine structure constant may be due to the non-minimal coupling of the electromagnetic field to a light scalar field.
This will induce the spatial variations of $\alpha$ because of the dynamical nature of the scalar field. In this paper, we studied the effects
of these spacetime variations on the CMB power spectra after the recombination. The time variation changes the redshift of the last scattering
surface and the amplitude of the power spectra slightly. The spatial variations induced extra anisotropies to the CMB temperature fluctuations
at the linear order like Sachs-Wolf effect. The shape of the spectra of polarizations and temperature-polarization cross spectrum are also modified
at higher orders. So in the studies of varying $\alpha$ theories, especially using high precision CMB data to constrain different models, the
effects from the spatial fluctuations of the fine structure constant should also be considered.
Furthermore these fluctuations induce additional $B$-polarizations similar to the effect of weak lensing. But
it is different from the effect originated from the scalar-photon anomaly coupling.
Hence the varying $\alpha$ theories provide a mechanism to produce CMB $B$ modes at later time, but these $B$ modes will mix with those from weak
lensing and it is not easy to distinguish them from each other. The more detailed results along this line will be published in the future \cite{wlz}.

\section{Acknowledgement}

We thank the referee for useful comments and suggestions.
This work is supported by Specialized Research Fund for the Doctoral Program of Higher Education
(SRFDP) under Grant No. 20090091120054.

{}

\end{document}